\documentclass[pre,twocolumn,groupedaddress,floatfix]{revtex4-1}
\usepackage{amssymb,amsmath}
\usepackage{graphicx}
\usepackage{subfigure}
\usepackage[english]{babel} 
\usepackage{float}
\usepackage{color}

\usepackage{lipsum}
\begin{document}
\newcommand{\be}{\begin{equation}}
\newcommand{\ee}{\end{equation}}
\newcommand{\rojo}[1]{\textcolor{red}{#1}}

\title{Transmission across a ribbon containing a square ${\cal PT}$ impurity}

\author{Cristian Mej\'{i}a-Cort\'{e}s$^{\dagger}$ and Mario I. Molina}
\email{mmolina@uchile.cl}
\affiliation{
$^{\dagger}$ Programa de F\'{i}sica, Facultad de Ciencias B\'{a}sicas,
Universidad del Atl\'{a}ntico, Puerto Colombia 081007, Colombia\\ 
$^{*}$ Departamento de F\'{\i}sica, Facultad de Ciencias, Universidad de Chile, Casilla 653, Santiago, Chile}

\date{\today }

\begin{abstract} 
We study the spectrum and transmission coefficient of plane waves propagating along square ribbons of varying widths, containing a square-shaped,  ${\cal PT}$-symmetric impurity region. We start with a zero-width ribbon (1D chain) and place a ${\cal PT}$ symmetric dimer. The spectrum is computed numerically and the instability gain is computed as a function of the gain/loss dimer strength. The transmission coefficient is obtained in closed form and examined as a function of wavevector and the gain/loss parameter. Next, we study a ribbon in a narrow ladder configuration containing a square ${\cal PT}$ impurity . As before, we compute the instability gain numerically and the transmission coefficient in closed form for the two possible input modes. Finally, we repeat the calculations for a wider ladder ribbon containing a Lieb-like impurity in a ${\cal PT}$ configuration.  For all cases and transmission channels, we obtain transmission divergences in wavevector-gain/loss parameter space, whose number increases with the width of the ribbon. 

\end{abstract}

\maketitle

\subsubsection{Introduction}
The topic of non-hermitian optics has attracted considerable interest in recent times\cite{bender,longhi,miri,ganainy}. It is based on the concept of optical materials that possess the so-called $\cal{PT}$ symmetry where, the real part of the dielectric function of an optical structure is symmetric in space, while its imaginary part is antisymmetric. This leads to a balance of gain and loss that allows novel ways to manipulate light. This has led to new phenomenology in 
optical wave scattering\cite{zyablovsky}, optical sensing\cite{sensing}, nonlinear optical processes\cite{nonlinear}, 
unidirectional reflectionless propagation\cite{feng,regensburger,lin}, single-mode lasing\cite{liu},  simultaneous lasing and coherent perfect absorption\cite{longhi2, chong}, among others.

On the other hand, the transport of excitations in low-dimensional lattices continues to be a subject of interest due to its technological applications to mesoscopic systems. For instance, in low-dimensional electronic systems, the transmission coefficient is directly linked to the electrical conductivity of the system\cite{landauer}. The great advances in nanotechnology have increased the interest in computing spectra, density of states, conductivity, etc for mesoscopic systems, in order to be able to manage and steer the magnetic, electrical, or optical response of these materials.

In this work, we study the spectrum and transmission of plane waves along ribbons of varying widths containing a square-shaped ${\cal PT}$-symmetric impurity region (Fig.1). Inside the impurity region 
we have sites with balanced gain and loss, which preserves $\cal{PT}$ symmetry.
Light propagating through the waveguide array can be described by the discrete Schr\"{o}dinger equation, whose stationary equation has the form
\be
-\lambda C_{\bf n}+ i g_{\bf n} + \sum_{\bf m} V_{\bf n m}\ C_{\bf m}=0\label{eq1}
\ee
where $\lambda$ is the propagation constant of the modes, $C_{n}$ is a mode amplitude, $\{g_{\bf n}\}$ is a spatial distribution of gain/loss propagation constants obeying the ${\cal PT}$ symmetry condition, $V_{\bf n m}$ is the coupling parameter, and the sum runs over all nearest neighbor sites. We will focus on the case $g_{\bf n}=\pm g$, and study how the physics of the system depends on the gain/loss parameter $g$.
\begin{figure}[t]
 \includegraphics[scale=0.3]{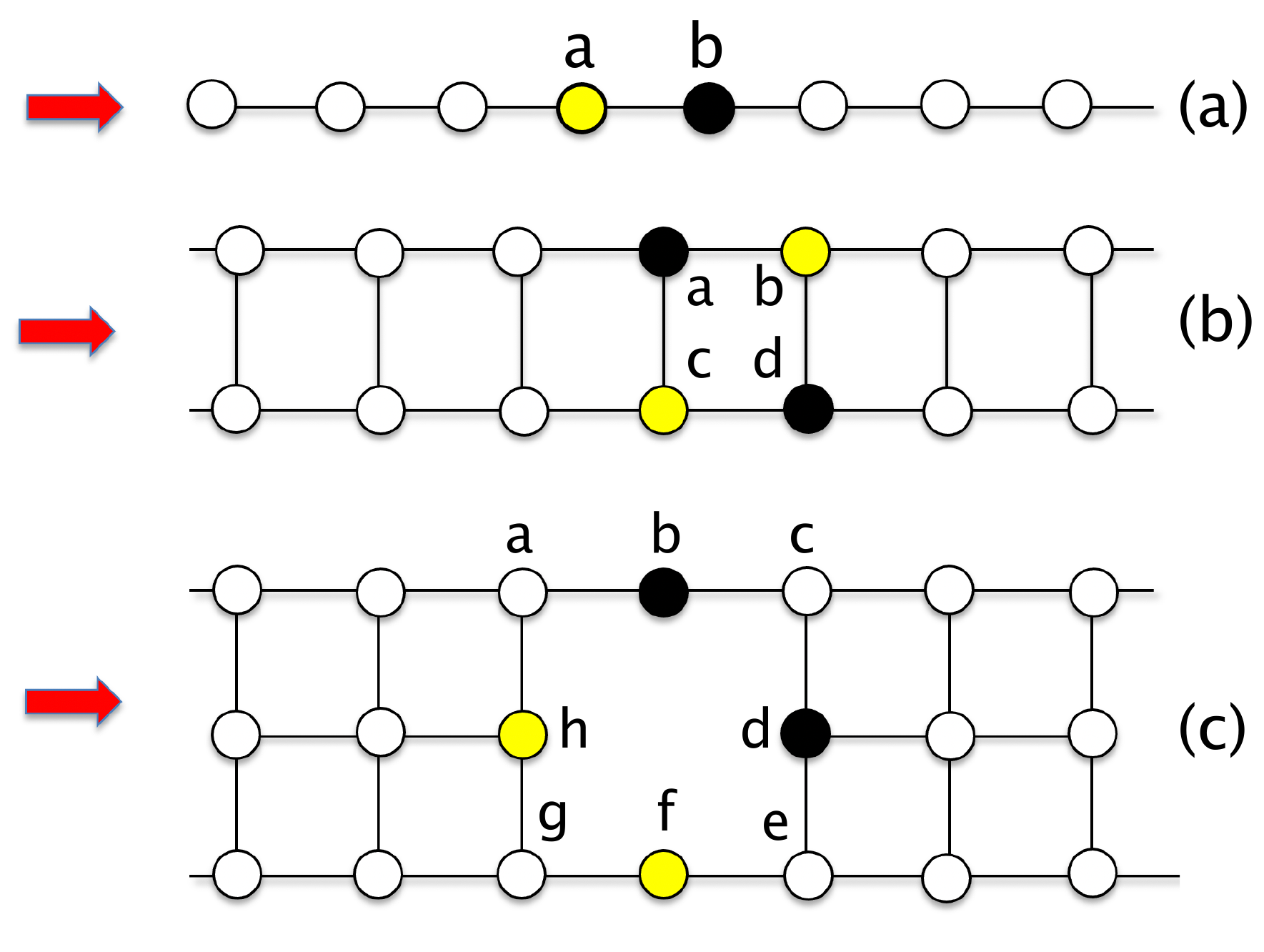}
  \caption{Ribbons of varying width containing a square-like ${\cal PT}$ impurity. (a)1D chain (b)Narrow ladder configuration (c) Wide ladder configuration. Black (yellow) colors denote sites with gain(loss), or viceversa.}  \label{fig0}
\end{figure}
What typically happens is that there is a critical value for the gain/loss strength below which the spectrum is real (symmetric phase), while above threshold, the system develops complex eigenvalues, bringing the system into the broken symmetry phase. 

The spatial extension of any given mode is monitored through the Participation Ratio $R$, defined as
\be
R = {( \sum_{n}|C_{n}|^2 )^2\over{\sum_{n} |C_{n}|^4}}.
\ee
For a completely localized mode, $R=1$, while for a completely delocalized one, $R=N$, where $N$ is the total number of lattice sites. One of the things we are interested in is the effect of the presence of $\cal{PT}$ effects on the average localization of the modes. Thus, for a given gain/loss parameter $g$ we compute the average $\langle R\rangle$ over all $N$ modes. Results of this procedure for our three lattices are shown in fig.2a. The symmetry $R(g)\rightarrow R(-g)$ originates from the symmetry 
$\{\lambda,g,C_{n}\} \Leftrightarrow \{-\lambda, -g, (-1)^n C_{n}\}$ of Eq.(\ref{eq1}). For all three cases, we see a delocalization regime in the vicinity of $g=0$.  Outside this small $g$ regime, the three $\langle R\rangle$ curves decrease to a value that remains constant as $g$ increases.  It is interesting to note that, in the wide ladder case, $\langle R\rangle$ is substantially smaller than for the other two cases at small $g$ values, and larger outside the small $g$ regime. This could be due to the fact that even in the absence of $\cal{PT}$ effects ($g=0$), the wide ribbon still gives rise to a vacancy impurity. This impurity causes some localization of its own, thus decreasing $\langle R\rangle$. This $g=0$ impurity is absent in the other two cases. On the other hand, at larger $g$ values, the higher value in $\langle R\rangle$ for the wide ribbon case could be explained as a simple geometric effect: In the wide lattice there are more sites around a given one for a mode to `expand into' and thus, decreasing the localizing tendency of the impurity. 
Summarizing, we could say that the presence of $\cal{PT}$ produces a tendency toward localization of the modes, for gain/loss parameter values greater than a critical one. 

The boundary between the symmetric and broken phase can be monitored through the instability rate 
\be
\gamma = \mbox{max}\{|\mbox{Im}[\lambda]|\}_{\lambda},
\ee
where $\lambda$ is an eigenvalue. Inside the symmetric phase $\gamma=0$, while at the broken phase, $\gamma>0$. Now we take a chain of length $N$ and, for a given $g$, compute all eigenvalues $\lambda$. From this, we compute the instability rate. 
Results are shown in Fig.2b for each of the three lattices. We notice that, as expected, there is a region of low $g$ values where $\gamma=0$, followed by a monotonic increase as $g$ is augmented.

\subsubsection{The 1D transmission case}
In this case the impurity region is modeled as a ${\cal PT}$ dimer embedded in a periodic chain (Fig.1a). Without loss of generality we place the dimer at sites $n=0$ and $n=1$, where all $g_{\bf n}=0$ except for $g_{0}=-g_{1}=g$, where $g$ is the gain/loss parameter. The embedded dimer in a linear chain has been treated before\cite{kivshar}, but we redo some calculations (in a slightly different form) for the sake of completeness. The stationary equations read
\be
-\lambda C_{n}+i g (\delta_{n 1}-\delta_{n 0}) C_{n}+V(C_{n+1}+C_{n-1})=0.
\ee
Let us look now at the transmission problem for the 1D case (Fig.1a). To the left of the dimer the mode 
has the form $R_{0} e^{i k n} + R e^{-i k n}$, where $R_{0}$ is the incident amplitude, $R$ is the reflected amplitude, and $k$ is the wavevector, while at the right of the dimer, we have a transmission amplitude $T e^{i k n}$. The stationary equations read
\begin{eqnarray}
& &(-\lambda - i g) C_{0}+V( R_{0} e^{-i k} + R e^{i k} + C_{1})=0\nonumber\\
& &(-\lambda + i g) C_{1}+V(C_{0}+T e^{2 i k})=0\nonumber\\
& &-\lambda T e^{2 i k} + V(T e^{3 i k}+C_{1})=0\nonumber\\
& &-\lambda (R_{0} e^{-i k}+ R e^{i k})+V(R_{0} e^{-2 i k}+ R e^{2 i k}+C_{0})=0,\nonumber\\ 
\end{eqnarray}
together with $\lambda = 2 V \cos(k)$. We define the transmission coefficient across the dimer ${\cal PT}$ defect as $t=|T/R_{0}|^2$, obtaining
\be
t = {4 V^4 \sin(k)^2\over{g^4 \cos(k)^2+(g^2-2 V^2)^2 \sin(k)^2}}\label{eq4}
\ee
and is shown in Fig.2c as a 3D phase diagram in wavevector-gain/loss space. From Eq.(\ref{eq4}) we find four points at which $t$ diverges: $k=\pm \pi/2, g = \pm \sqrt{2} V$. The transmission is also invariant under the exchange of gain and loss sites, $t(g)=t(-g)$.
\begin{figure}[b]
 \includegraphics[scale=0.3]{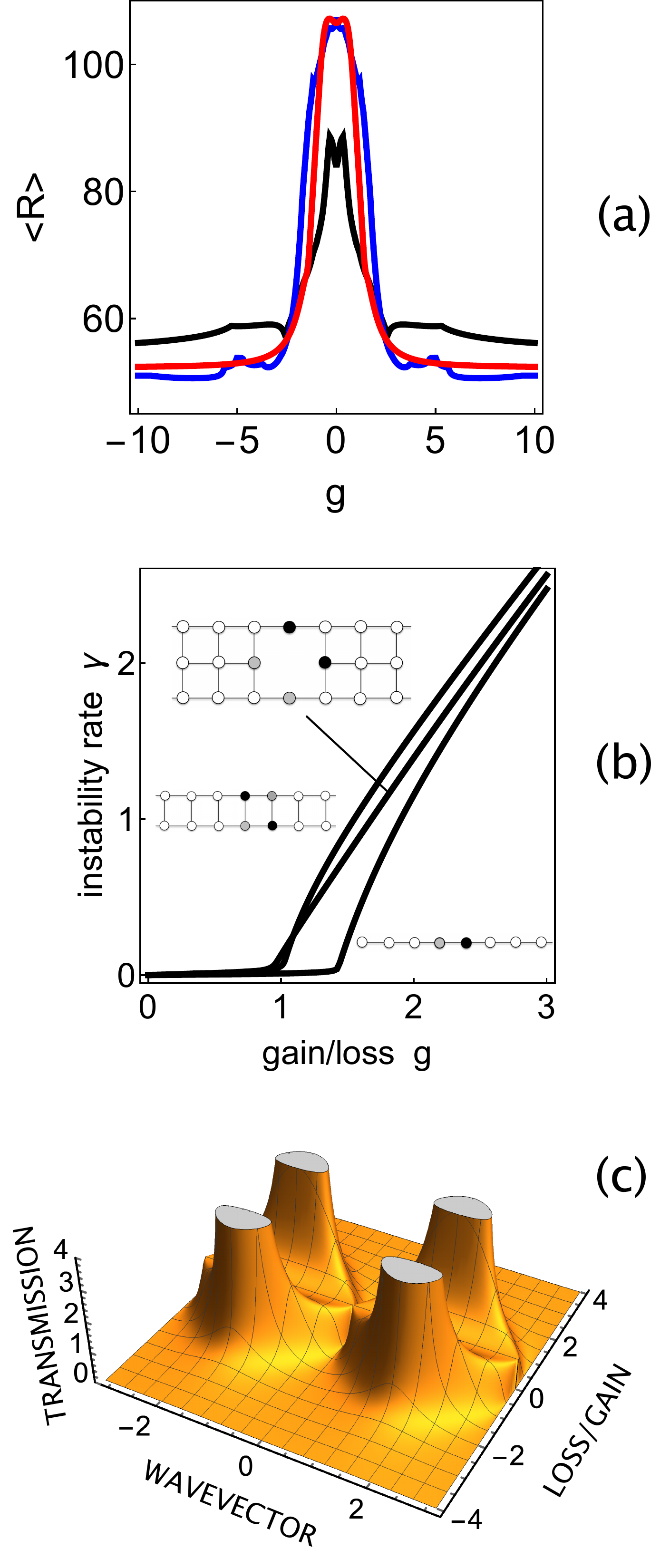}
  \caption{ ${\cal PT}$ square-like impurities embedded in a quasi-1D chain. (a) Average participation ratio vs gain/loss parameter, for the 1D (red), narrow ladder (blue), and wide ladder (black) cases ($N=159$). 
  (b) Instability rate as a function of gain/loss parameter ($N=300$). (c) The transmission coefficient of plane waves across the ${\cal PT}$ dimer, as a function of wavevector and gain/loss parameter. At $k=\pm \pi/2, g=\pm \sqrt{2}$, the transmission diverges.}  \label{fig1}
\end{figure}
As can be seen in Fig.2c, the region on the $g$ axis where the transmission is greater than unity is concentrated on a finite region around $g=0$. Outside this small region, the transmission falls abruptly. This region is bounded by the curve $g=\pm 2 V \sin(k)$. The values of transmission greater than unity can be traced to the presence of the impurity as a source or sink, and the current is therefore, not conserved. This is true even in the $\cal{PT}$ symmetric phase. The instability rate (Fig.2b) is nearly zero for $g\lesssim 1.4$, followed by a monotonic increase.

\subsubsection{The narrow ladder configuration}
The ladder is composed of two 1D chains forming side rails joined by rungs and containing a ${\cal PT}$ square impurity (Fig.1b). 
A related ladder system consisting of an infinitely extended distribution of $\cal{PT}$ symmetric site energies, and containing four nonlinear impurities in a square configuration has also been considered recently\cite{malomed}. All nearest neighbor sites are coupled with coupling coefficient $V$. We denote by $C_{n}^{u}$ ($C_{n}^{d}$) the amplitude at the upper (lower) side rail joined by the $n$th rung. Without loss of generality we place the ${\cal PT}$ impurity sites on $(0,u), (0,d), (1,u)$ and $(1,d)$. 
The stationary equations are
\begin{eqnarray}
-\lambda& C_{n}^u& + V (C_{n}^l+C_{n+1}^u+C_{n-1}^u)=0\nonumber\\
-\lambda& C_{n}^l& + V (C_{n}^u+C_{n+1}^l+C_{n-1}^l)=0\label{eq5}
\end{eqnarray}
outside the square impurity, while at the impurities the equations are
\begin{eqnarray}
-\lambda C_{0}^u& + & i g\; C_{0}^u+V (C_{0}^l+C_{1}^u+C_{-1}^u)=0\nonumber\\
-\lambda C_{0}^l& + & i g\; C_{0}^l+V (C_{0}^u+C_{1}^d+C_{-1}^l)=0\nonumber\\
-\lambda C_{1}^u& - & i g\; C_{1}^u+V (C_{1}^l+C_{2}^u+C_{0}^u)=0\nonumber\\
-\lambda C_{1}^l& - & i g\; C_{1}^l+V (C_{1}^u+C_{2}^l+C_{0}^l)=0.
\end{eqnarray}
In the absence of the impurity region, the stationary equations (\ref{eq5}) give rise to two bands $\lambda^{+} = 2 V \cos(k) + V$, with eigenmode $(C_{n}^u,C_{n}^d)=(1/\sqrt{2})(1,1) e^{i k n}$ (mode channel $1$) and  $\lambda^{-} = 2 V \cos(k) - V$, with eigenmode $(C_{n}^u,C_{n}^d)=(1/\sqrt{2})(1,-1) e^{i k n}$ (mode channel $2$).


For the transmission problem, we take $A\times (1,1) e^{i k n}$ as the input beam, i.e., along mode channel $1$. After colliding with the ${\cal PT}$ square impurity, we have two reflection modes:$ R_{1}\times (1,1) e^{-i k n}$ and $ R_{2}\times (1,-1) e^{-i k n}$, and two transmission modes: $T_{1}\times (1,1) e^{i k n}$ and $T_{2}\times (1,-1) e^{i k n}$. We denote by $a,b,c$ and $d$ the amplitudes at the impurity sites with general site energies $\epsilon_{a}, \epsilon_{b}, \epsilon_{c}$ and $\epsilon_{d}$. The stationary equations read
\begin{eqnarray}
(-\lambda +\epsilon_{a}) a+ (b + c + I e^{-i k}+(R_{a}+R_{b})\;e^{i k})=0\ \ \ \ \ \\
(-\lambda +\epsilon_{b}) b+ V(a + d +(T_{1}+T_{2})e^{2 i k})=0\ \ \ \ \ \\
(-\lambda + \epsilon_{c}) c+V(a + d + I e^{-i k}+(R_{1}-R_{2}) e^{i k}) =0\ \ \ \ \ \\
(-\lambda + \epsilon_{d}) d+V(b+c+(T_{1}-T_{2})e^{2 i k}) =0\ \ \ \ \ \\
-\lambda (I e^{-i k}+(R_{1}+R_{2})e^{i k}) + V(a + I e^{-2 i k}+\ \ \ \ \ \nonumber\\
+(R_{1}+R_{2})e^{2 i k}+ I e^{-i k}+(R_{1}-R_{2})e^{i k})=0\ \ \ \ \ \\
-\lambda (I e^{-ik}+(R_{1}-R_{2})e^{ik})+V(c + I e^{-2 i k}+\ \ \ \ \ \nonumber\\
+(R_{1}-R_{2}) e^{2 i k}+I e^{-i k}+(R_{1}+R_{2})e^{ik})=0\ \ \ \ \ \\
-\lambda (T_{1}+T_{2}) e^{2 i k}+V(b + (T_{1}+T_{2}) e^{3 i k}+\ \ \ \ \ \nonumber\\
+(T_{1}-T_{2}) e^{2 i k})=0\ \ \ \ \ \\
-\lambda (T_{1}-T_{2}) e^{2 i k}+V(d + (T_{1}-T_{2}) e^{3 i k}+\ \ \ \ \ \nonumber\\
+(T_{1}+T_{2}) e^{2 i k})=0\ \ \ \ \ 
\end{eqnarray}
where $\epsilon_{a},\epsilon_{b},\epsilon_{c}$ and $\epsilon_{d}$ are the complex site energies at the impurity sites. In what follows we take $\epsilon_{a}= i g, \epsilon_{b}= -i g, \epsilon_{c}= -i g$ and $\epsilon_{d}=i g$, i.e., a ${\cal PT}$ configuration, where $g$ is the gain/loss strength. The transmission coefficient along channel $(1,1)$ is given by $t_{1}\equiv |T_{1}/I|^2$, while the transmission through channel $(1,-1)$ is given by $t_{2}\equiv |T_{2}/I|^2$. After solving the above linear system, we obtain
\be
t_{1}(k,g)=\left|{\mbox{num}_1(k,g)\over{\mbox{den}_1(k,g)}}\right|^2, 
\ee
where,
\begin{gather}
\mbox{num}_1(k,g)=(-1+e^{2 i k})(1+8 e^{i k}+16 e^{5 i k}+4 e^{6 i k}+\nonumber\\
4 e^{4 i k}(7+g^2)+4 e^{3 i k}(8+g^2)
+e^{2 i k}(23+g^2)),
\end{gather}
\begin{gather}
\mbox{den}_1(k,g)= 1+8 e^{i k}-16 e^{7 i k}-4 e^{8 i k}+\nonumber\\
+12 e^{3 i k}(2+g^2)+2 e^{2 i k}(11+g^2)+4 e^{6 i k}(-6+g^4)\nonumber\\
+4 e^{5 i k}(-4+3 g^2+g^4)+e^{4 i k}(5 + 22 g^2+g^4).
\end{gather}

On the other hand, the transmission through channel $(1,-1)$ is given by $t_{2}\equiv |T_{2}/I|^2$, where 
\be
t_{2}(k,g)=\left|{\mbox{num}_2(k,g)\over{\mbox{den}_2(k,g)}}\right|^2, 
\ee
with,
\begin{gather}
\mbox{num}_2(k,g)= 2 g (1+e^{-i k})^2(-1+e^{2 i k}),
\end{gather}
\begin{gather}
\mbox{den}_2(k,g)= 5+8 e^{-3 ik}-16 e^{3 i k}+e^{-4 i k}-4 e^{4 i k}+\nonumber\\
22 g^2+g^4+12 e^{-i k}(2+g^2)+2 e^{-2 i k}(11+g^2)+\nonumber\\
+4 e^{2 i k}(-6+g^2)+ 4 e^{i k}(-4+3 g^2+g^4).
\end{gather}
Results for $t_{1}(k,g)$ and $t_{2}(k,g)$ are shown in Fig.3. 
\begin{figure}[t]
\includegraphics[scale=0.31]{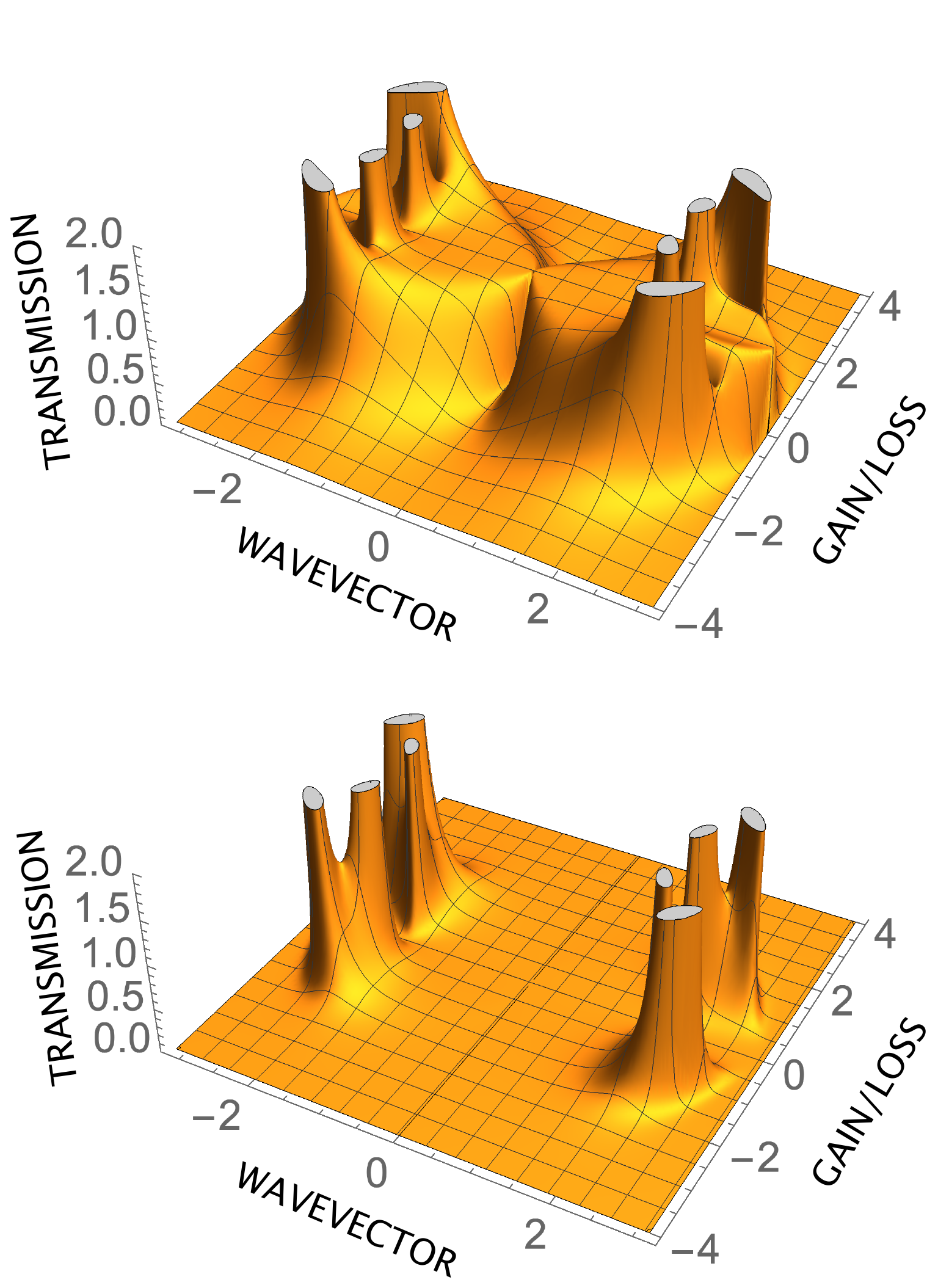}
 \caption{Narrow ladder: Transmission coefficient $t_{1}(k,g)$ of mode channel $1$ (top) and $t_{2}(k,g)$ of mode channel $2$  (bottom) for an input wave of the form $(1,1) e^{i k n}$.}  \label{fig2}
\end{figure}
We notice the symmetry $t_{1,2}(k, -g)=t_{1,2}(k,g)$ and $t_{1,2}(-k, g)=t_{1,2}(k,g)$ and the observation that, as in the 1D case, the region in wavevector-gain/loss space where $t_{1,2}$ is greater than unity, is restricted to a finite vicinity around $(k,g)=(0,0)$ and is composed of eight separated `peaks' of infinite height. Away from the peak cluster, the transmission decreases monotonically with the gain/loss parameter $g$.	The peaks of $t_{1,2}(k,g)$ are located at $(\pm 2.419, \pm 1.323),(\pm 1.939,\pm 0.599)$.

\subsubsection{The wide ladder configuration} 
Finally, we consider the ribbon of width $2$ containing a Lieb-like impurity (Fig.1c). The procedure for computing the eigenvalues, the instability rate,  and the several transmissions is the same as the one employed in the previous sections.  We obtain an instability rate that is qualitatively similar to the previous case (Fig.3a), with a nearly zero amplitude for $\gamma \lessapprox 1$ and rising monotonically from there.
In the absence of the impurity, we have three bands: $\lambda^{(1)}=2 V \cos(k)+\sqrt{2} V, \lambda^{(2)}=2 V \cos(k)$ and $\lambda^{(3)}=2 V \cos(k)-\sqrt{2} V$, with corresponding modes: $C^{(1)}_{n}\sim (1,\sqrt{2},1) e^{i k n}$, $C^{(2)}_{n}\sim (1,0,-1)e^{i k n}$ and $C^{(3)}_{n}\sim (1,-\sqrt{2},1) e^{i k n}$. If we take the input beam as belonging to mode channel $1$, we have the input state $A (1,\sqrt{2},1) e^{i k n}$. Upon reflection, we have three possible reflection channels: $R_{1} (1,\sqrt{2},1)e^{-i k n}$, $R_{2} (1,0,-1)e^{-i k n}$ and $R_{3} (1,-\sqrt{2},1)e^{-i k n}$. Similarly, we have three transmission channels: $T_{1} (1,\sqrt{2},1) e^{i k n}$, $T_{2} (1,0,-1) e^{i k n}$ and $T_{3} (1,-\sqrt{2},1) e^{i k n}$. 
\begin{figure}[t]
 \includegraphics[scale=0.3]{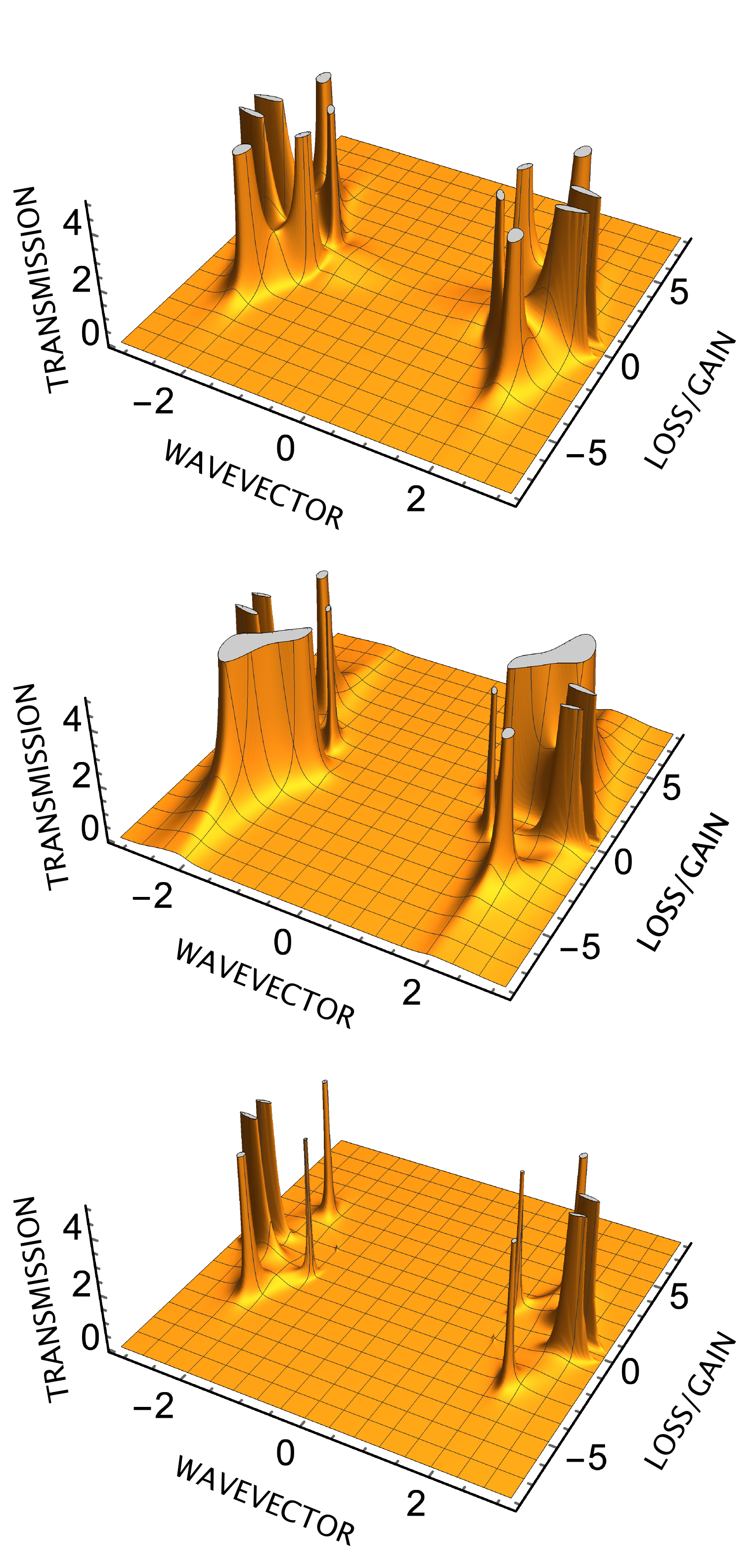}
  \caption{Wide ladder: Transmission coefficients as a function of wavevector and gain/loss, along mode channel $1$ (top), mode channel $2$ (middle) and mode channel $3$ (bottom) for an input wave of the form $(1,\sqrt{2},1) e^{i k n}$.}  \label{fig3}
\end{figure}
We denote by $a, b, c, d, e, f, g$ and $h$ the amplitudes at the impurity sites with site energies 
$\epsilon_{a}$, $\epsilon_{b}$, $\epsilon_{c}$, $\epsilon_{d}$, $\epsilon_{e}$, $\epsilon_{f}$, 
$\epsilon_{g}$, and $\epsilon_{h}$. For the transmission problem, the modes have the form
\begin{gather}
(C_{n}^u, C_{n}^m, C_{n}^d)=A (1,\sqrt{2},1) e^{i k n}\hspace{0.3cm}\mbox{$n<0$\ \ \  {\bf input wave}}\nonumber\\
(C_{n}^u, C_{n}^m, C_{n}^d)=R_{1} (1,\sqrt{2},1) e^{-i k n} + R_{2} (1,0,-1) e^{-i k n}+\nonumber\\
           + R_{3} (1,-\sqrt{2},1) e^{-i k n}\hspace{0.6cm}\mbox{$n<0$ {\bf reflected}}\nonumber\\
(C_{n}^u, C_{n}^m, C_{n}^d)=T_{1}(1,\sqrt{2},1) e^{i k n}+T_{2} (1,0,-1) e^{i k n}+\nonumber\\
+T_{3} (1,-\sqrt{2},1) e^{i k n}\hspace{0.3cm}\ \ \ \mbox{$n>2$ {\bf transmitted}}\nonumber
\end{gather}
where $C_{n}^u, C_{n}^m$ and $C_{n}^d$ refer to the mode amplitude at the nth rung and upper, middle and lower rail, respectively. This nomenclature applies to those sites outside the $\cal{PT}$ impurity region.

The relevant stationary equations are those at the impurity sites and the sites immediately next to them. The resulting $14$ stationary equations read:
\begin{gather}
(-\lambda +\epsilon_{a}) a+V(b + h + A e^{-i k}+(R_{1}+R_{2}+R_{3})\;e^{i k})=0\nonumber\\
(-\lambda +\epsilon_{b}) b+V(a+c)=0\nonumber\\
(-\lambda +\epsilon_{c}) c+V(b+d+(T_{1}+T_{2}+T_{3}) \exp(3 i k))=0\nonumber\\
(-\lambda +\epsilon_{d}) d+V(c+e+\sqrt{2}(T_{1}-T_{3})\exp(3 i k))=0\nonumber\\
(-\lambda +\epsilon_{e}) e+V(d+f+(T_{1}-T_{2}+T_{3})\exp(3 i k)=0\nonumber\\
(-\lambda +\epsilon_{f}) f+V(e+g)=0\nonumber\\
(-\lambda +\epsilon_{g}) g+V(f+h+A e^{-i k}+(R_{1}-R_{2}+R_{3}) e^{i k})=0\nonumber\\
(-\lambda +\epsilon_{h}) h+V(g+a+\sqrt{2}(A e^{-i k}+(R_{1}-R_{3}) e^{i k}))=0\nonumber\\
-\lambda(T_{1} + T_{2} + T_{3}) e^{3 i k} + 
  V (c + (T_{1} + T_{2} + T_{3}) e^{4 i k} +\nonumber\\ 
     +\sqrt{2} (T_{1} - T_{3}) e^{3 i k} )=0\nonumber\\
     -\lambda \sqrt{2} (T_{1} - T_{3}) e^{3 i k} + 
   V ( d + \sqrt{2} (T_{1} - T_{3}) e^{4 i k} + \nonumber\\
   +(T_{1} - T_{2} + T_{3}) e^{3 i k} + (T_{1} + T_{2} + T_{3}) e^{3 i k} )=0\nonumber\\        
        -\lambda(T_{1} - T_{2} + T_{3}) e^{3 i k} + 
  V ( e + (T_{1} - T_{2} + T_{3}) e^{4 i k} + \nonumber\\
     +\sqrt{2} (T_{1} - T_{3}) e^{3 i k} )=0\nonumber\\
-\lambda (A e^{-i k} + (R_{1} - R_{2} + R_{3}) e^{i k}) + 
  V ( g + A e^{-2 i k} + 
  \nonumber\\
 +(R_{1} - R_{2} + R_{3}) e^{2 i k} + 
     \sqrt{2} (A e^{-i k} + (R_{1} - R_{3}) e^{i k }) )=0\nonumber\\
     -\lambda \sqrt{2} (A e^{-i k} + (R_{1} - R_{3}) e^{i k}) + 
  V (h + \sqrt{2} (A e^{-2 i k} + \nonumber\\
  +(R_{1} - R_{3}) e^{2 i k}) + 
     A e^{-i k} + (R_{1} + R_{2} + R_{3}) e^{i k} + 
     A e^{-i k} + \nonumber\\
     +(R_{1} - R_{2} + R_{3}) e^{i k} )=0\nonumber\\
  -\lambda ( A e^{-i k} + (R_{1} + R_{2} + R_{3}) e^{i k} ) + 
  V ( a + A e^{-2 i k} + \nonumber\\
  +(R_{1} + R_{2} + R_{3}) e^{2 i k} + 
     \sqrt{2} (A e^{-i k} + (R_{1} - R_{3}) e^{i k} ))=0
        \end{gather}

with $\lambda=2 V \cos(k) + \sqrt{2} V$.

After solving the linear system of equations in closed form (solutions for $R_{i}, T_{i}$ and for the amplitudes at each impurity site are not shown here for they are rather unwieldy), and after setting up the $\cal{PT}$ distribution for the impurities shown in Fig.1, we obtain all reflection and transmission amplitudes. 
The transmission on each mode channel is defined as
\be
t_{i} = {|T_{i}|^2\over{|A|^2}}
\ee
Results for $t_{1}, t_{2}$ and $t_{3}$ are shown in Fig.4. As before, we immediately notice the existence of `islands' in the wavevector--gain/loss phase space where the transmission on any channel is greater than unity and can in fact, diverge.

For all three cases, the effect of $\cal{PT}$ symmetry is the same: The creation of a localized region in $g$ space around $g=0$ where the transmission can be substantially greater than unity. These regions are composed of smaller `islands' inside which the transmission diverges at isolated points.

\subsubsection{Conclusions}
We have investigated the instability rate, mode localization, and transmission coefficient of plane waves across a $\cal{PT}$-symmetric impurity region in the shape of a `square' impurity,  embedded in a quasi-1D ribbon. For the three ribbons examined, the instability rate is zero for gain/loss parameters below threshold, for all cases examined, and increases monotonically beyond that. The average mode size measured by the participation ratio shows an abrupt transition from a state of large  mode size to a phase with lower mode size that remains constant over further increase in gain/loss parameter.  We could say that the presence of ${\cal PT}$ produces a tendency toward localization of the modes, for gain/loss parameter values greater than a critical one. 
Finally, the transmission coefficient across all impurity regions for all channels, shows localized regions in wavevector-gain/loss space where the transmission is greater than one, diverging at some isolated $g$ values. This is unsurprising, given that the presence of gain/loss effects implies a lack of current conservation, in principle. The great amplification of the transmission coefficient for specific $(k,g)$ suggests an amplifying or filtering mechanism in the optics context\cite{suchkov,lindquist}. 
The symmetry $k\rightarrow -k$ ensures the reciprocity of the system and is in agreement with standard results\cite{rayleigh}. Further examination of other diverse $\cal{PT}$ impurity configurations (not presented here) yields similar qualitative results, suggesting the generality of the observed results for ribbons containing a $\cal{PT}$ impurity region.

A possible extension of this work would be to consider a nonlinear PT impurity region. As a first approximation, one could focus on the simplest 1D case with two nonlinear impurities in the absence of PT effects. In that case, a previous study\cite{molina-tsironis 1993} has shown that the presence of nonlinearity increases the value of the transmission coefficient of plane waves, as well as creating a nonlinearity window where bistable effects can occur. When ${\cal PT}$ is included in the form of a dimer ${\cal PT}$ on the already nonlinear sites, it is observed the onset of an asymmetric transmission of plane waves across the dimer; the magnitude of such asymmetry increases with the value of the gain/loss parameter\cite{ambroise}. Also, the plane wave states evidence a generic instability. 
We believe these results could still hold in the limit of weak nonlinearity and thin ribbons. Work in this direction is already under way and results will be published elsewhere.
\vspace{0.5cm}

\noindent
{\em Acknowledgments}
\vspace{0.5cm}

This work was supported by Fondecyt Grant 1200120.\\


\begin{thebibliography}{99}

\bibitem{bender}
C. M. Bender and S. Boettcher, Real spectra in non-hermitian
Hamiltonians having PT symmetry, Phys. Rev. Lett. {\bf 80}, 5243
(1998).

\bibitem{longhi}
S. Longhi, Parity-time symmetry meets photonics: A new twist
in non-Hermitian optics, Europhys. Lett. {\bf 120}, 64001 (2017).

\bibitem{miri}
M.-A. Miri and A. Al\'{u}, Exceptional points in optics and photonics, Science {\bf 363}, 7709 (2019).

\bibitem{ganainy}
L. Feng, R. El-Ganainy, and L. Ge, Non-Hermitian photonics
based on parity-time symmetry, Nat. Photon. {\bf 11}, 752 (2017).

\bibitem{zyablovsky}
A A Zyablovsky, A P Vinogradov, A A Pukhov, A V Dorofeenko, A A Lisyansky, ${\cal PT}$-symmetry in optics,  Physics Uspekhi {\bf 57}, 1063 (2014).

\bibitem{sensing}
De Carlo, Martino, Francesco De Leonardis, Richard A. Soref, Luigi Colatorti, and Vittorio M. N. Passaro, Non-Hermitian Sensing in Photonics and Electronics: A Review, Sensors {\bf 22}, 3977 (2022).

\bibitem{nonlinear}
Vladimir V. Konotop, Jianke Yang, and Dmitry A. Zezyulin,
Nonlinear waves in ${\cal PT}$-symmetric systems,
Rev. Mod. Phys. {\bf 88}, 035002 (2016).

\bibitem{feng}
L. Feng, Y.-L. Xu, W. S. Fegadolli, M.-H. Lu, J. E. B.
Oliveira, V. R. Almeida, Y.-F. Chen, and A. Scherer, Experimental demonstration of a unidirectional reflectionless paritytime metamaterial at optical frequencies, Nat. Mater. {\bf 12}, 108(2013).

\bibitem{regensburger}
A. Regensburger, C. Bersch, M.-A. Miri, G. Onishchukov, D. N.
Christodoulides, and U. Peschel, Parity-time synthetic photonic
lattices, Nature (London) {\bf 488}, 167 (2012).

\bibitem{lin}
Z. Lin, H. Ramezani, T. Eichelkraut, T. Kottos, H. Cao,
and D. N. Christodoulides, Unidirectional Invisibility Induced
by PT-Symmetric Periodic Structures, Phys. Rev. Lett. {\bf 106},
213901 (2011).

\bibitem{liu}
W. Liu, M. Li, R. S. Guzzon, E. J. Norberg, J. S. Parker, M.
Lu, L. A. Coldren, and J. Yao, An integrated parity-time symmetric wavelength-tunable single-mode microring laser, Nat. Commun. {\bf 8}, 15389 (2017).

\bibitem{longhi2}
S. Longhi, PT-symmetric laser absorber, Phys. Rev. A {\bf 82},
031801(R) (2010). 

\bibitem{chong}
Y. D. Chong, L. Ge, and A. D. Stone, PT -Symmetry Breaking and Laser-Absorber Modes in Optical Scattering Systems, Phys. Rev. Lett. {\bf 106}, 093902 (2011).

\bibitem{landauer}
R. Landauer, ``Spatial Variation of Currents and Fields Due to Localized Scatterers in Metallic Conduction'',  IBM J. Res. Dev. {\bf 1}, 223 (1957).

\bibitem{kivshar}
Sergey V. Dmitriev, Sergey V. Suchkov, Andrey A. Sukhorukov, and Yuri S. Kivshar, Scattering of linear and nonlinear waves in a waveguide array with a PT -symmetric defect, Phys. Rev. A {\bf 84}, 013833 (2011).

\bibitem{malomed}
J. D'Ambroise, S. Lepri, B.A. Malomed, and P.G. Kevrekidis,
$\cal{PT}$-symmetric ladders with a scattering core, Phys. Lett. A {\bf 378}, 2824 (2014). 

\bibitem{suchkov}
S. V. Suchkov, D. I. Borisov, A. A. Sukhorukov, Yu. S. Kivshar, Signal manipulation with a PT-symmetric coupler embedded into an array of optical waveguides, Lett. Mater. {\bf 4}, 222 (2014).

\bibitem{lindquist}
B. Lindquist, Design of filters in a one-dimensional tight-binding system, Phys. Rev. E {\bf 63}, 56605 (2001).

\bibitem{rayleigh}
J. Rayleigh, The theory of sound (Dover publications, New York, 1945).

\bibitem{molina-tsironis 1993}
M. I. Molina and G. P. Tsironis, Nonlinear impurities in a linear chain, 
Phys. Rev. B {\bf 47}, 15330 (1993).

\bibitem{ambroise}
J. D'Ambroise, P. G. Kevrekidis, and S. Lepri,
Asymmetric wave propagation through nonlinear ${\cal PT}$-symmetric oligomers,
J. Phys. A: Math. Theor. {\bf 45}, 444012 (2012).






\end{thebibliography}
\end{document}